\documentclass[twocolumn,preprintnumbers,amsmath,amssymb,superscriptaddress]{revtex4}
\usepackage{graphicx}
\usepackage{dcolumn}
\usepackage{bm}
\usepackage{braket}
\usepackage{amsmath}

\newcommand{\comment}[1]{}

\newcommand{\yb}{$^{174}$Yb }

\newcommand{\g}{^{1}S_{0}}

\newcommand{\tpl}{^{3}P_{1}}

\newcommand{\omrec}{\omega_{\rm rec}}

\usepackage{color}

\begin{document}
	
	\title{Bloch-bands Picture for Light Pulse Atom Diffraction and Interferometry}

	\author{Daniel Gochnauer}
	\author{Katherine E. McAlpine}
	\author{Benjamin Plotkin-Swing}
	\affiliation{Department of Physics, University of Washington, Seattle Washington, 98195, USA}
	\author{Alan O. Jamison}
	\affiliation{Department of Physics and MIT-Harvard Center for Ultracold Atoms, Research Laboratory of Electronics, MIT, Cambridge, Massachusetts 02139, USA}
	\author{Subhadeep Gupta}
	\affiliation{Department of Physics, University of Washington, Seattle Washington, 98195, USA}
	
	\date{\today}
	\begin{abstract}
	We apply a Bloch-bands approach to the analysis of pulsed optical standing wave diffractive elements in optics and interferometry with ultracold atoms. We verify our method by comparison to a series of experiments with Bose-Einstein condensates. The approach provides accurate Rabi frequencies for diffraction pulses and is particularly useful for the analysis and control of diffraction phases, an important systematic effect in precision atom interferometry. Utilizing this picture, we also demonstrate a method to determine atomic band structure in an optical lattice through a measurement of phase shifts in an atomic contrast interferometer.  	
	\end{abstract}
	\maketitle	
	
\section{Introduction}

The motion of electrons in the periodic potential of ionic crystals is addressable by the celebrated Bloch solutions which give rise to band structure \cite{bloc28, ashc76}. Periodic potentials are also common in the arena of ultracold atoms where gases trapped in optical lattices can serve as a test-bed for questions in many-body physics \cite{bloc08,bloc12,kuhr16}. These scenarios are amenable to the same band structure approach. 

Pulsed optical lattices are in common use as diffractive elements in atom optics and interferometry \cite{tino14} for diverse applications such as inertial sensing \cite{mcgu02,durf06,dutt16,geig11} and for tests of fundamental physics such as the equivalence principle \cite{fray04,schl14, tara14} and quantum electrodynamics \cite{bouc11,park18}. The atom-optics element of choice for beamsplitters and mirrors --- Bragg diffraction --- is traditionally analyzed using the two-state Rabi solution which predicts oscillatory behavior \cite{gilt95,gupt01}. To address the regime when the two-state approximation is invalidated for sufficiently short pulses, a host of numerical work has been performed \cite{durr99,horn99,mull08b} with the limiting case of Kapitza-Dirac diffraction allowing an analytic solution \cite{gupt01}. 

In this work we apply the Bloch-bands approach to atom optics through the performance and analysis of a series of standing-wave diffraction and interferometry experiments with Bose-Einstein condensates. Our results impact three key directions. First, we experimentally demonstrate the equivalence between the band gap and the frequency for Bragg pendell{\"o}sung, and obtain accurate values distinct from the results of a commonly-used formula for Rabi frequency in Bragg diffraction. Second, we exploit the Bloch-bands approach for direct visualization and systematic analysis of diffraction phase effects, and provide useful methods for their suppression in precision atom interferometry. Finally, we invert the approach and determine atomic band structure in a periodic potential from measurements of phase shifts in an atom interferometer, thus introducing a new method for analyzing arbitrary periodic optical potentials.

\section{Atom optics in the Bloch-bands picture}

\begin{figure}[h]
		\center
		\includegraphics[width=0.5\textwidth]{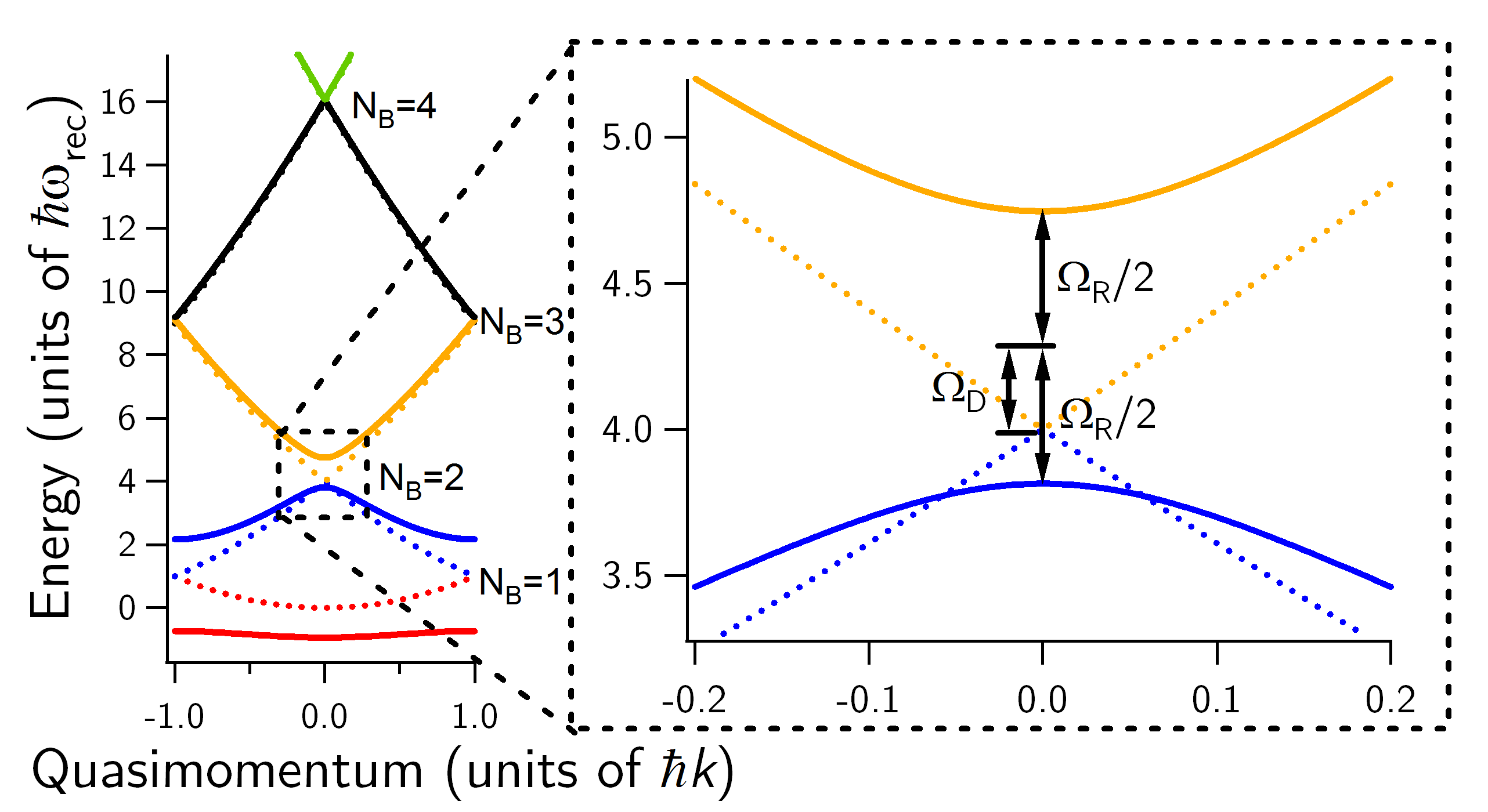}
		\caption{(Color online) Atomic energy bands in quasimomentum space for a sinusoidal optical lattice. Solid (dashed) lines are calculated with a depth, $U_0$, of $6\hbar \omrec$ ($0\hbar \omrec$). The first through fourth order Bragg transitions are indicated by $N_B=1$ to 4. Inset shows a close-up for $N_B=2$. $\Omega_R$ and $\Omega_D$ correspond to the frequencies of Rabi flopping and diffraction phase evolution respectively.}
		\label{fig:Gochnauer_BandPicture_Figure1}
\end{figure}

Our analysis is based on the Bloch solutions for a neutral atom interacting with the one-dimensional sinusoidal potential of an optical standing wave and is related to earlier theoretical work \cite{cham01,buch03}. In accord with parameters used in typical experiments, we work in a regime where the large one-photon detuning allows adiabatic elimination of the excited internal state. The atom-light interaction then reduces to a conservative (AC Stark shift) potential imposed on the atoms, which is proportional to the optical intensity \cite{gupt01}. We calculate the Bloch energy bands (Fig.\ref{fig:Gochnauer_BandPicture_Figure1}) by numerically diagonalizing the single-particle Hamiltonian for the potential $U = U_0 \sin^2(2kx)$. The energy and momentum are normalized to the recoil energy $E_{\rm rec}=\hbar^2 k^2/2m$ and recoil momentum $p_{\rm rec}=\hbar k$ respectively, where $\pi/k$ is the spatial periodicity of the lattice.

In a Brillouin-zone picture, the lattice opens up an avoided crossing at every intersection of free-particle energy levels, each of  which can be identified with a Bragg diffraction process and is characterized by an energy gap which increases monotonically with $U_0$. This band gap is equivalent to $\hbar \Omega_R$, where $\Omega_R$ is the Rabi frequency for oscillations between the two Bragg-coupled states. In addition, there is also an energy shift $\hbar \Omega_D$ of the mean energy of the coupled states away from the original (unperturbed) crossing point. Both $\Omega_R$ and $\Omega_D$ can be seen as arising from ``level repulsion'' in second-order perturbation (see Fig.\ref{fig:Gochnauer_BandPicture_Figure1} inset).

We now turn to determining the Bragg diffraction amplitude and phase in this picture. We explicitly consider an $N_B^{\rm th}$ order Bragg transition which can be seen as a $2N_B$ photon process connecting states $\ket{-N_B\hbar k}$ and $\ket{+N_B\hbar k}$. Here even (odd) $N_B$ corresponds to a crossing at the center (edge) of the Brillouin zone. In the band picture, the Bragg process is the behavior of an initially free particle state which is loaded into the lattice at the $N_B^{\rm th}$ avoided crossing as an equal superposition of the $N_B^{\rm th}$ and $(N_B-1)^{\rm th}$ excited bands (Bloch states). During the $2N_B$ photon pulse the population in the $\ket{\pm N_B\hbar k}$ states oscillate sinusoidally and out of phase with each other at angular frequency $\Omega_R(t)$. Since each of the $\ket{\pm N_B\hbar k}$ states spends equal time in each band, we may evaluate the corresponding phase by integrating the average energy of the two participating Bloch states, characterized by $\Omega_D$, over the duration of the pulse. Within an atom interferometer, diffraction pulses are frequently applied to a superposition of free particle states separated in momenta by multiples of $2\hbar k$, corresponding to different interferometer paths. The different $\Omega_D(t)$ for different paths during these processes can then lead to an observable interferometer phase shift called the {\it diffraction phase}.   

In order to apply the Bloch-bands picture to pulsed standing waves, the time-dependence of $U_0(t)$ must preserve the two-state nature of the Bragg process. Practically, the desire for high diffraction efficiency means experiments work in exactly this regime, showing the suitability of the Bloch-bands approach.

For a time-varying standing-wave amplitude $U_0(t)$, adiabaticity mandates $\frac{1}{U_0}\frac{\partial U_0}{\partial t} \ll \Delta E/\hbar $ where $\Delta E$ is the energy separation from the eigenstate nearest to our two states of interest. Applying this criterion to the rise and fall times of a smooth (e.g., Gaussian) pulse shape with width $\tau$, resonant with an $N_B^{\rm th}$ order Bragg process, we arrive at $\tau \gg \frac{1}{4 N_B \omrec}$ where $\omrec=E_{\rm rec}/\hbar$ is the recoil frequency. We recognize this inequality to be equivalent to being in the Bragg regime of diffraction where states other than $\ket{\pm N_B\hbar k}$ are not populated.

We compare the Bloch band picture to both full numerical time evolution of the Schr{\"o}dinger equation and experiments. In Bloch picture simulations, we numerically solve for the band structure at many different lattice depths. We then use these saved band structures to numerically integrate $\Omega_R$ and $\Omega_D$ for any given diffraction pulse profile to obtain population fraction and diffraction phase predictions. With a single set of band structures, the effects of any pulse shape or duration may be quickly calculated, allowing for rapid prototyping of experimental sequences. These simulations enforce adiabaticity by assuming the atoms' wavefunctions remain confined to the two bands corresponding to their free-space momenta. By contrast, full numerical evolution solves the time-dependent Schr{\"o}dinger equation over a large basis of momentum states and uses the time evolution operator to extract population transfer and diffraction phase information (see Appendix B). A full time evolution must be calculated for every pulse shape or duration considered. Comparing to full time evolution allows us to quantify the non-adiabatic effects due to other bands missing from the adiabatic Bloch picture.

 We test and verify the validity of the Bloch-bands picture of atoms optics in the $\tau \gtrsim  \frac{1}{4 N_B \omrec}$ ``quasi-adiabatic'' regime (see Appendix A) and stay within the observed validity range for all the experimental work presented in this paper. Even when other states are negligibly populated, their presence has a significant effect for typical experimental parameters on both the splitting $\Omega_R$ and the shift $\Omega_D$ of the two coupled states. We now examine these effects individually. 

\section{Accurate Rabi Frequencies for Bragg Diffraction}

\begin{figure}[h]
		\center
		\includegraphics[width=0.5\textwidth]{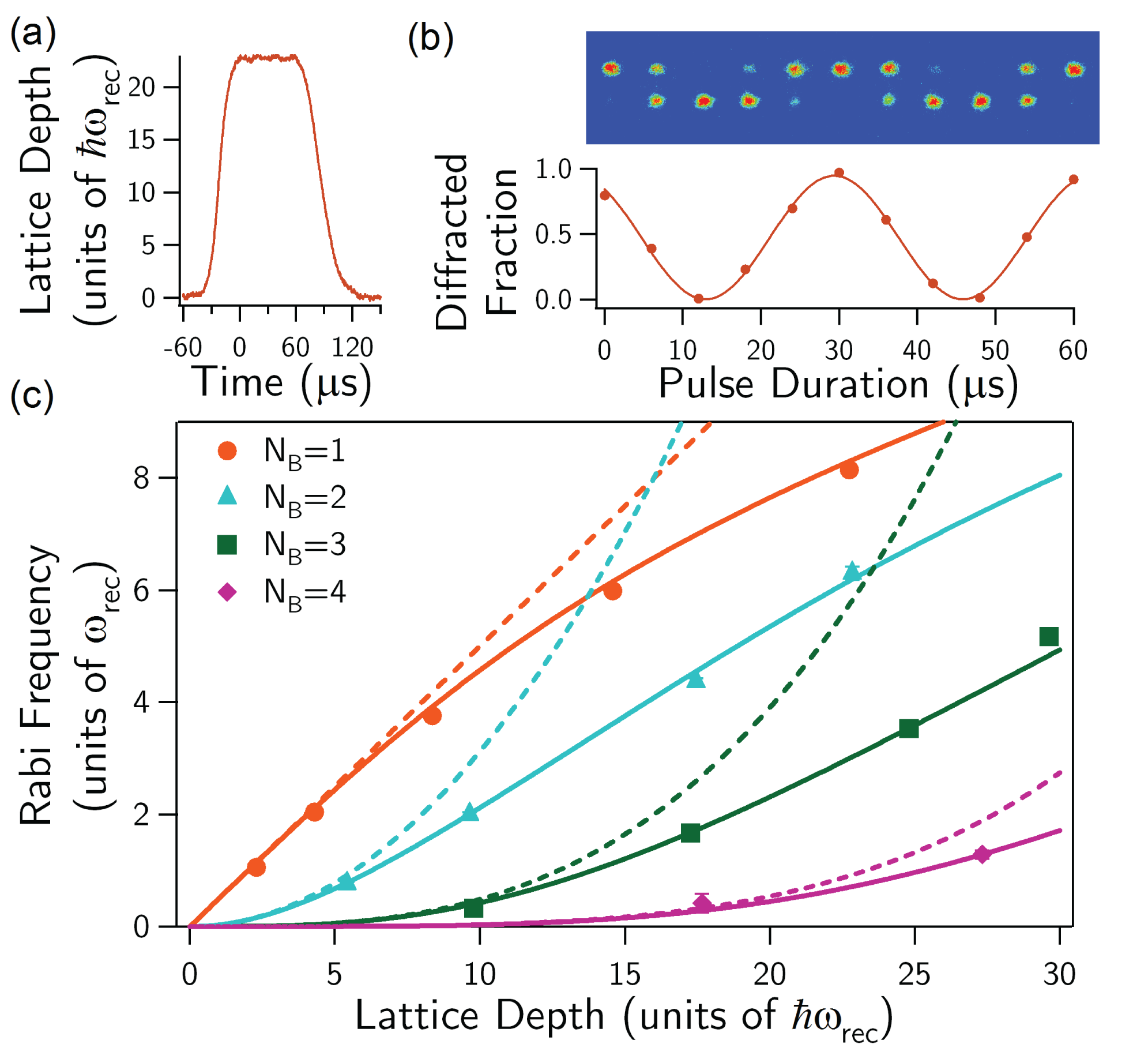}
		\caption{(Color online) Measurement of band gaps (Rabi frequencies). (a) Sample pulse profile. (b) Rabi oscillations for $N_B=1$ in a $22.7\,\hbar \omrec$ lattice; upper panel shows corresponding sequence of time-of-flight absorption images. The pulse duration is defined as the extent of the intermediate flat region of the pulse profile. The diffracted fraction for zero pulse duration is due to the pulse rise and fall. (c) Measured Rabi frequency for various lattice depths and for $N_B=1,2,3,4$ (filled circles).  Solid (dashed) lines are Bloch-bands (RHS of Eqn.\ref{eqn:RabiEqn}) predictions.}
		\label{fig:Gochnauer_BandPicture_Figure2}
\end{figure}

We first report on our measurements of the Rabi frequency for various Bragg diffraction orders using a BEC atom source. Our results experimentally establish the Bloch-bands picture for atom optics and reveal the shortcomings of a commonly used result in the field (see Fig.\ref{fig:Gochnauer_BandPicture_Figure2}).

The experiments reported in this work were carried out with Bose-Einstein condensates (BECs) of $10^5$ ytterbium (\yb\!) atoms. We prepared the BEC in a $532\,$nm crossed-beam optical dipole trap and released them from the confinement after reducing the mean trap frequency to $\overline{\omega} = 2\pi \times 63\,$Hz \cite{plot18}. Upon release, the atoms are given $2\,$ms to expand before they encounter a diffraction pulse. The diffraction pulses are formed by a pair of horizontal, counterpropagating beams with a waist of $1.8\,$mm, blue detuned from the $556\,$nm intercombination line $(\g \rightarrow \tpl)$ by $+3500\Gamma$, where $\Gamma = 2\pi \times 182\,$kHz is the natural linewidth. The relative detuning between the two beams is controlled by direct digital synthesis electronics at the sub-Hz level. The size of the cloud during all the atom optics experiments is 34 $\mu$m, i.e., far less than the size of the diffraction beams. The depth of the optical lattice formed by these beams was calibrated with Kapitza-Dirac diffraction \cite{gupt01}. This method provided a depth calibration accurate at the $\pm2\%$ level.

We applied diffraction pulses with temporal intensity profiles consisting of Gaussian rise and fall $1/e$ times of $\tau_{\rm 1/e}=27\,\mu$s satisfying $4\omrec \tau_{\rm 1/e}=2.5$, with an intermediate flat profile of variable extent (Fig.\ref{fig:Gochnauer_BandPicture_Figure2}(a)). The relative detuning $\delta$ of the lattice beams was set to match the Bragg resonance condition $\delta = 4 N_B \omrec$. The population in each of the two coupled states was monitored by time-of-flight absorption imaging (Fig.\ref{fig:Gochnauer_BandPicture_Figure2}(b), upper panel). The fractional population in the final state oscillates as $P(t)=\sin^2[\frac{1}{2}\int_{t_0}^t\Omega_R^{(2N_B)}(t'){\rm d}t']$ where $\Omega_R^{(2N_B)}$ is the Rabi frequency for an $N_B^{\rm th}$ order Bragg process (Fig.\ref{fig:Gochnauer_BandPicture_Figure2}(b)). As shown in Fig.\ref{fig:Gochnauer_BandPicture_Figure2}(c), the measured $\Omega_R^{(2N_B)}$ is in good agreement with the Bloch-bands calculation. 

Fig.\ref{fig:Gochnauer_BandPicture_Figure2}(c) also demonstrates the inadequacy of a commonly used \cite{gupt01,jans07,mazz15,huqi17} generalized Rabi frequency formula first derived in \cite{gilt95}, given by the RHS of the inequality below:
	\begin{equation}
		\Omega_R^{(2N_B)} < \frac {[\omega_R]^{2N_B}}{2^{4N_B-3}[(N_B-1)!]^2\Delta^{N_B}\omrec^{N_B-1}}
		\label{eqn:RabiEqn}
	\end{equation}
Here $\omega_R$ is the single photon Rabi frequency and $\Delta$ is the detuning from the excited state. The RHS is a perturbative result and therefore breaks down at large lattice depth, deviating significantly from the measured values. It is important to note that the standard pulse parameters used in Bragg diffraction experiments are comfortably outside this perturbative regime, stemming from the favoring
of short pulses in experiments in order to minimize state manipulation time in comparison to the longer interferometer interaction times. 

\begin{figure}[h]
	\center
	\includegraphics[width=0.5\textwidth]{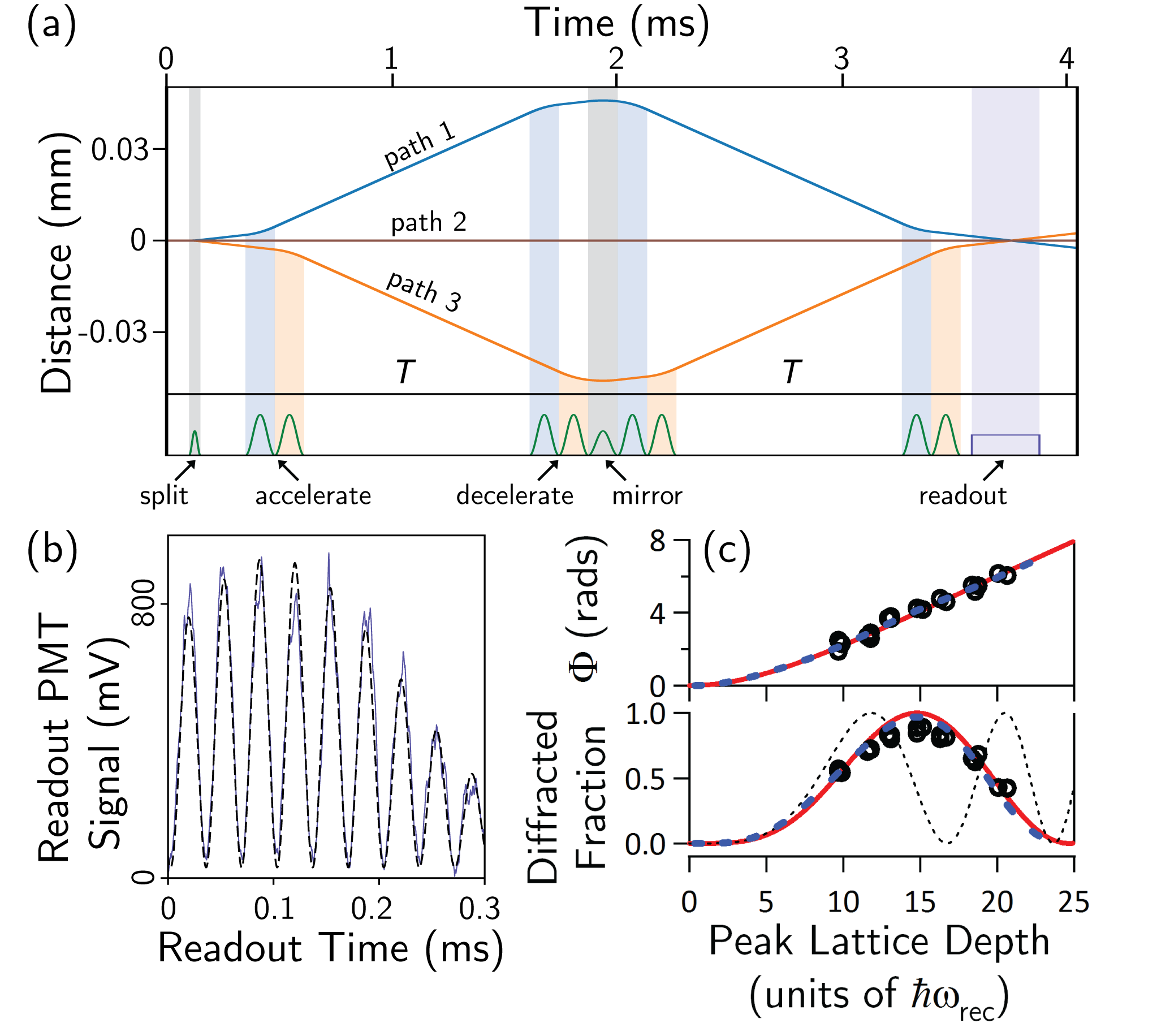}
	\caption{(a) Space-time diagram for a CI with up to $16\hbar k$ momentum separation between the outer paths and free evolution time, $T=1\,$ms. (b) Representative readout signal for the CI (20 shot average) together with fitted sinusoid with a Gaussian envelope. (c) Varying the peak lattice depth of the mirror pulse changes both the diffracted fraction and the CI phase $\Phi$. The solid red lines show the corresponding Bloch band analysis predictions while the blue dashed lines show the full numerical integration theory, with both methods displaying good agreement with the data (open circles). The black dotted line is the prediction from the RHS of Eqn.\ref{eqn:RabiEqn}.}
	\label{fig:Gochnauer_BandPicture_Figure3}
\end{figure}

\section{Application to Diffraction Phases in Atom Interferometry}

The Bloch-bands picture allows a straightforward understanding and assessment of diffraction phases in atom interferometry. In addition to the $\hbar \Omega_R$ band gap, the perturbation of the lattice produces a shift $\hbar \Omega_D$ of the mean energy of the two coupled states (Fig.\ref{fig:Gochnauer_BandPicture_Figure1}(b)). In the presence of a lattice of depth $U_0$, a particular path within an interferometer is characterized by a particular band number and quasimomentum $q$, accumulating an additional phase during a diffraction process:
\begin{equation}
	\Phi_D = \frac{1}{\hbar}\int_{\rm pulse}({\bar{E}}(q,U_0)-E_f(q))dt
	\label{eqn:dphase}
\end{equation}
where $E_f$ is the free particle energy. For an interferometer path at Bragg resonance during the pulse, ${\bar {E}}(t) = E_f+\hbar \Omega_D(t)$; however, away from a Bragg resonance it is the energy of the band the path is loaded into. Bragg diffraction within an atom interferometer thus results in differential phase shifts between interferometer paths and can lead to an overall diffraction phase, with important ramifications for precision measurements \cite{buch03,jami14,este15}. 

We perform our experimental work on diffraction phases in a three-path contrast interferometer (CI) (see Fig.\ref{fig:Gochnauer_BandPicture_Figure3}(a)) with a \yb BEC source \cite{jami14,plot18}. After release of our BEC from the trap, atoms are first placed in an equal superposition of three momentum states ($\ket{+2\hbar k}$, $\ket{0\hbar k}$, $\ket{-2\hbar k}$) using a short standing-wave light pulse operating in the Kapitza-Dirac regime \cite{gupt01}. The three parts of the wavefunction separate for time $T$ after which the outer paths have their momenta reversed by a second-order Bragg $\pi$-pulse. The three paths are again spatially overlapped after an additional time $T$. The contrast of the resulting matterwave interference pattern is measured as the Bragg reflection signal of a traveling-wave light pulse. This readout signal (Fig.\ref{fig:Gochnauer_BandPicture_Figure3}(b)) has the oscillating form:
\begin{equation}
	S(t)=C(t)\cos^2(\frac{\phi_1(t)+\phi_3(t)}{2} - \phi_2(t))
	\label{eqn:CIsignal}
\end{equation}
where $\phi_i$ is the phase accrued by path $i$ and $C(t)$ is an envelope function determined by the coherence time of the condensate source \cite{plot18}. $S(t)$ oscillates at $8\omrec$ and is sensitive to the kinetic energy differences between the interfering paths and thus to the photon recoil frequency, and can therefore be used to precisely measure the fine structure constant \cite{plot18}. We fit such signals with the expression $C(t_r){\rm cos}^2(4\omrec t_r+\Phi)+S_0$ where $t_r$ is the time from the start of the readout pulse and $S_0$ is a vertical offset. The momentum separation between outer paths during free evolution can be increased to $n\hbar k$ by the insertion of acceleration pulses (Fig.\ref{fig:Gochnauer_BandPicture_Figure3}(a) shows $n=16$) with the resulting CI phase $\Phi=\frac{1}{2}n^2\omrec T+\Phi_{\rm offset}$. Diffraction phase effects are contained in $\Phi_{\rm offset}$, which we study by keeping $T$ fixed and monitoring $\Phi$ for varying diffraction pulse parameters. All acceleration pulses used in this work are either second or third order Bragg pulses with Gaussian-shaped temporal profiles and are incorporated into the theoretical model as integrals over a time-varying $U_0(t)$ in Eqn.\ref{eqn:dphase}.

\begin{figure}[h]
		\center
		\includegraphics[width=0.5\textwidth]{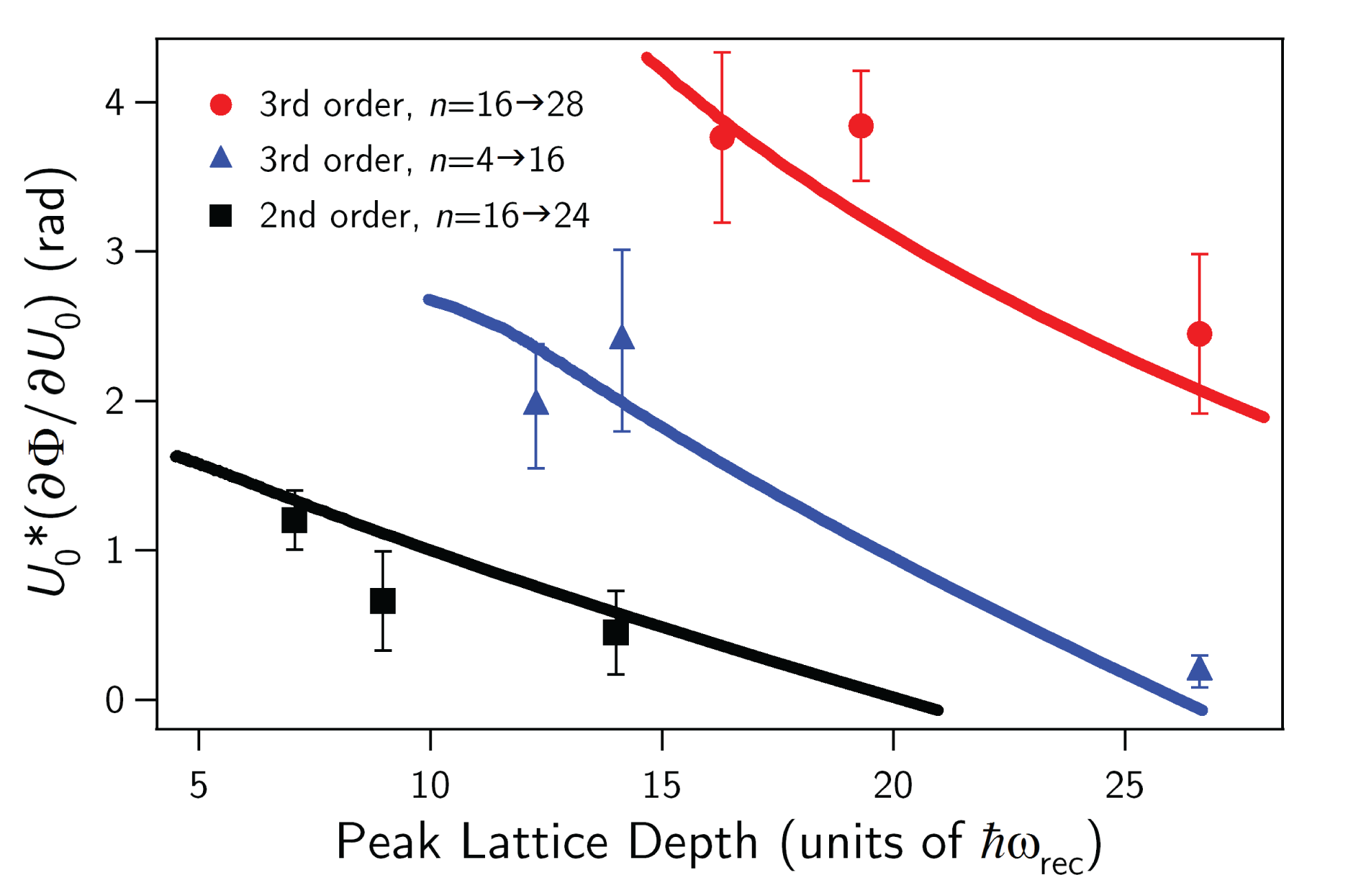}
		\caption{(Color online) Suppression of diffraction phase effects with pulse intensity. Measurements of $U_0 \partial \Phi/\partial U_0$ at the $\pi-$point for different Bragg acceleration pulses are in agreement with the Bloch-bands calculation (curves). The pulses accelerated the outer paths from $\ket{\pm2\hbar k}$ to $\ket{\pm8\hbar k}$ (blue triangles), $\ket{\pm8\hbar k}$ to $\ket{\pm12\hbar k}$ (black squares), and $\ket{\pm8\hbar k}$ to $\ket{\pm14\hbar k}$ (red circles).}
		\label{fig:Gochnauer_BandPicture_Figure4}
\end{figure}

As shown in Fig.\ref{fig:Gochnauer_BandPicture_Figure3}(c) for the Bragg (second order) mirror pulse, both the observed phase and the population oscillation are captured well by the Bloch-bands model. While paths 1 and 3 acquire diffraction phase according to $N_B=2$, path 2 remains in the lowest band. The observed sub-unity diffraction efficiency at the $\pi$-point (where the $\pi$-pulse condition is met) is due to the small but finite velocity width of the sample, which is not included in the model.

Since interferometry experiments are sensitive to diffraction phase through intensity noise (i.e., shot-to-shot variations in $U_0$), we characterize its effect using the measured quantity $U_0 \partial \Phi /\partial U_0$ at operating conditions, which for the CI geometry are $\pi$-pulses for both mirror and acceleration optics. This quantity serves as a good figure of merit for the typical situation where the uncertainty in depth scales with $U_0$, since it normalizes the phase fluctuations at each depth. From datasets similar to Fig.\ref{fig:Gochnauer_BandPicture_Figure3}(c), we determine the slopes at these $\pi-$points for several different pulse parameters of $2^{\rm nd}$ and $3^{\rm rd}$ order Bragg acceleration pulses (see Fig.\ref{fig:Gochnauer_BandPicture_Figure4}) and find good agreement with the Bloch-bands theory. When the momentum separation between the paths is large compared to recoil (i.e., multi-band separation), this quantity becomes negligible except for the path(s) undergoing the Bragg transition. Our analysis of diffraction phases in the Bloch-bands picture shows that $\Omega_D/\Omega_R$ monotically decreases with lattice depth in Bragg diffraction for $N_B>1$ (see Fig.\ref{fig:Gochnauer_BandPicture_Figure7}(b) in Appendix C). This is the reason for the observed decrease of $U_0 \partial \Phi / \partial U_0$ with increasing $U_0$ in Fig.\ref{fig:Gochnauer_BandPicture_Figure4}.

This result points to an important consequence for precision interferometry: for high-order Bragg diffraction as commonly used for large momentum separation interferometers \cite{chio11, plot18}, diffraction phase effects are minimized by operating at as high a lattice depth as possible, as long as additional states are not populated by the process. This can be understood as the slowed growth of $\Omega_D$ with $U_0$ from the level repulsion of higher energy states (see Appendix C). Another result of our analysis is that the diffraction phase can be significantly modified by the presence of other interferometer paths in a nearby band (see Eqn.\ref{eqn:CIsignal}) as can be seen in the difference between the red and blue data points in Fig.\ref{fig:Gochnauer_BandPicture_Figure4}. This method can be used to greatly suppress the diffraction phase effect (see blue data point at 27 recoils in Fig.\ref{fig:Gochnauer_BandPicture_Figure4}). We can also see that certain interferometer geometries are immune to diffraction phases from the symmetry of pulse application, e.g., the symmetric Mach-Zehnder. However interferometers that measure the recoil phase accrued between paths are generally sensitive to diffraction phases \cite{jami14,este15,park18,plot18}. 

\section{Determining Band Structure from Interferometer Phase}

While the earlier discussion has been mainly focussed on the application of the Bloch-bands picture at avoided crossings, the picture applies equally well to all other points in quasimomentum space. This approach to atom diffraction and interferometry can thus naturally lend itself as a tool to determine band structure due to some unknown periodic potential. The transient presence of some unknown optical lattice manifests as an interferometric phase shift with different interferometer paths evolving phase according to the band number and quasimomentum into which they map. By varying these quantities with the atom optical elements of the interferometer, the complete band structure due to the unknown potential can be determined.

\begin{figure}[h]
	\center
	\includegraphics[width=0.5\textwidth]{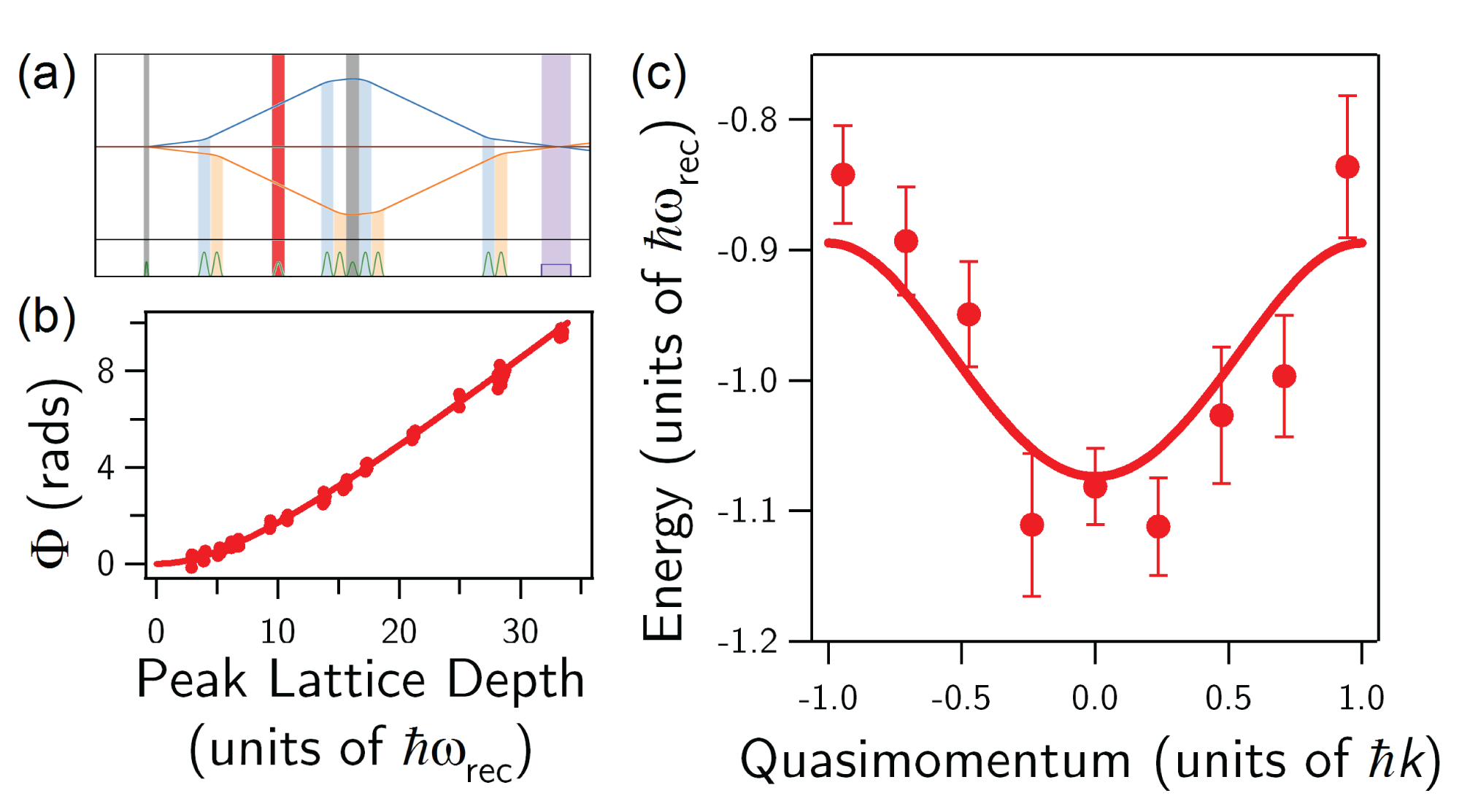}
	\caption{(a) Space-time diagram for band structure measurement with a modified $n=16$ CI. This is identical to Fig.\ref{fig:Gochnauer_BandPicture_Figure3}(a) except for the additional pulsed lattice (dark red stripe) which imparts the band structure to be determined. (b) CI phase at $q=0$ for various lattice strengths. (c) The ground band dispersion in a sinusoidal lattice from diffraction phase measurements. The solid line corresponds to the theoretical dispersion for a lattice depth of $6.5\hbar \omrec$.}
	\label{fig:Gochnauer_BandPicture_Figure5}
\end{figure}
	
A clean implementation of this tool is furnished in the CI, modified as shown in Fig.\ref{fig:Gochnauer_BandPicture_Figure5}(a). For demonstration purposes we determine the ground band structure in a sinusoidal optical lattice with Gaussian temporal shape and $4\omrec \tau_{\rm 1/e}=2.5$. The value of quasimomentum $q$ in the standing-wave frame is chosen by precisely controlling the relative detuning of the counterpropagating lattice beams in the lab frame. During the pulse, the middle path is in the bottom band (band 0) and the outer paths are in bands 7 and 8. Diffraction phase accrued by the outer paths is negligible compared to that accrued by the middle path. Fig.\ref{fig:Gochnauer_BandPicture_Figure5}(b) shows the mesaured CI phase at $q=0$ for various lattice depths, in good agreement with the band theory prediction \cite{footrabi}. As we vary $q$ at a fixed depth, the measured CI phase converted to an energy shows the characteristic ground band dispersion (Fig.\ref{fig:Gochnauer_BandPicture_Figure5}(c)).

The data presented in Fig.\ref{fig:Gochnauer_BandPicture_Figure5}(c) was obtained through a series of experiments in which a non-diffracting lattice is applied within the interferometer at a time highlighted (red stripe) in Fig.\ref{fig:Gochnauer_BandPicture_Figure5}(a). Because the relative detuning of the two lattice beams was chosen such that the middle interferometer path loaded into the bottom band, the other paths --- having been accelerated --- were loaded into much higher bands and thus contributed negligibly to the total CI phase. The quasimomentum $q$ in the standing-wave frame could then be varied and any phase shift would be an observation of the lattice energy dispersion. The adiabaticity criterion prevents turning on a lattice at $q$ near the Brillouin zone edge without loading into a superposition of the bottom band and the first excited band. To measure the energy shift at a particular $q$ value in only the bottom band, we developed the following procedure: First we adiabatically turned on a lattice at $q=0$, reaching a depth of $6.5\hbar \omrec$ over 150 $\mu$s with a cubic spline temporal shape. Next we linearly ramped the relative detuning between diffraction beams for 56 $\mu$s until we reached $q=-0.95\hbar k$. Then we swept the relative detuning at the same rate in the opposite direction, stopping at the desired $q$ for 100 $\mu$s, and eventually reaching $q=+0.95\hbar k$. Finally, the relative detuning was brought back to zero and the lattice turned off with the same cubic spline shape.

This method thus allowed us to directly measure the diffraction phase $\Phi_D$ as a function of quasimomentum. In accord with Eqn.\ref{eqn:dphase}, we converted $\Phi_D$ to an energy shift and then added the calculated free-space energy at each $q$ to obtain the ground band dispersion. In the experimental sequence the diffraction phase acquired during each intensity ramp and frequency ramp was common for all experimental iterations, and resulted in a uniform offset phase in the experiment. This became an energy offset in the measured ground band dispersion, which we determined by fitting with one free parameter (the constant offset) to the calculated ground band energy dispersion for a $6.5\hbar \omrec$ depth lattice. The obtained ground band dispersion determined from the data in this way shows good agreement with the theoretical calculation (Fig.\ref{fig:Gochnauer_BandPicture_Figure5}(c)).

\section{Discussion and Conclusions}

We have investigated a Bloch-bands approach to analyzing atom diffraction and interferometry. Theoretical results for the amplitude and phase associated with standing wave diffraction show good agreement with measurements for arbitrary lattice strength. Significantly, our results span a range of atom optics parameters that extend beyond the weak lattice regime, and are thus of practical importance for current atom interferometry experiments. While analytic formulas for diffraction amplitude are known from earlier work \cite{gilt95,gupt01} and their inadequacy beyond the weak lattice limit recognized \cite{cham01,jans07}, our results constitute the first experimental study and its accurate analysis at arbitrary lattice depth. We have for the first time demonstrated the validity of a Bloch-bands approach to diffraction phases by direct comparison to interferometric phase shifts. Our work also points out general methods to control diffraction phases, an important systematic effect in precision measurements. All our results, presented as scaled to recoil frequency and momentum are generally valid for all atom diffraction and interferometry setups. Our interferometric method of band structure measurement is complementary to earlier methods \cite{klin10,geig18}, and while demonstrated here only for the ground band, can be extended to excited bands as well by loading path 2 into the desired band. Furthermore, the method can also be adapted to non-sinusoidal periodic potentials, as well as to higher-dimensional and time-dependent (e.g., Floquet) lattices \cite{fuji19}.

We thank Tahiyat Rahman for experimental assistance. This work was supported by NSF Grant No. PHY-1707575 and by the NASA Fundamental Physics Program. 

\appendix

\section{Adiabaticity Criterion}	
	
The potential from the time-varying diffraction pulse can be written as $U(x,t)=U_0(t){\rm sin}^2(2kx)$. From the adiabatic theorem of quantum mechanics, we can write down the adiabaticity criterion for a time-varying standing-wave amplitude $U_0(t)$ as:
\begin{equation}
	\frac{1}{U_0}\frac{\partial U_0}{\partial t} \ll \Delta E(q)/\hbar 
\end{equation}
where $\Delta E(q)$ is the energy separation from the eigenstate nearest to our two states of interest. For the typical experimental situation of a pulse with smooth rise and fall times with width $\tau$ resonant with an $N_B^{\rm th}$ order Bragg process, the adiabatic criterion gives  $\tau \gg \frac{1}{4 (N_B-1) \omrec} \approx \frac{1}{4 N_B \omrec}$ for $N_B>1$ and where $\omrec=E_{\rm rec}/\hbar$ is the recoil frequency. For $N_B=1$, $\tau \gg \frac{1}{8 \omrec}$. The adiabatic criterion then is equivalent to being in the Bragg regime of diffraction where states other than the two coupled ones are not populated. 

The Doppler width from the finite velocity spread $\Delta v$ of the atomic source introduces another important timescale and the corresponding mandate of $\tau \ll \frac{1}{k\Delta v}$ is met by using sub-recoil clouds of atoms. Additionally, practical considerations in atom optics applications put a premium on diffraction pulse times seeking to minimize this in favor of interferometer interaction times. It is thus critical to test the regime of validity of the Bloch-bands picture of atoms optics in the regime $\tau \gtrsim  \frac{1}{4 N_B \omrec}$, or a ``quasi-adiabatic'' regime. 
	
\begin{figure}[t]
	\center
	\includegraphics[width=0.48\textwidth]{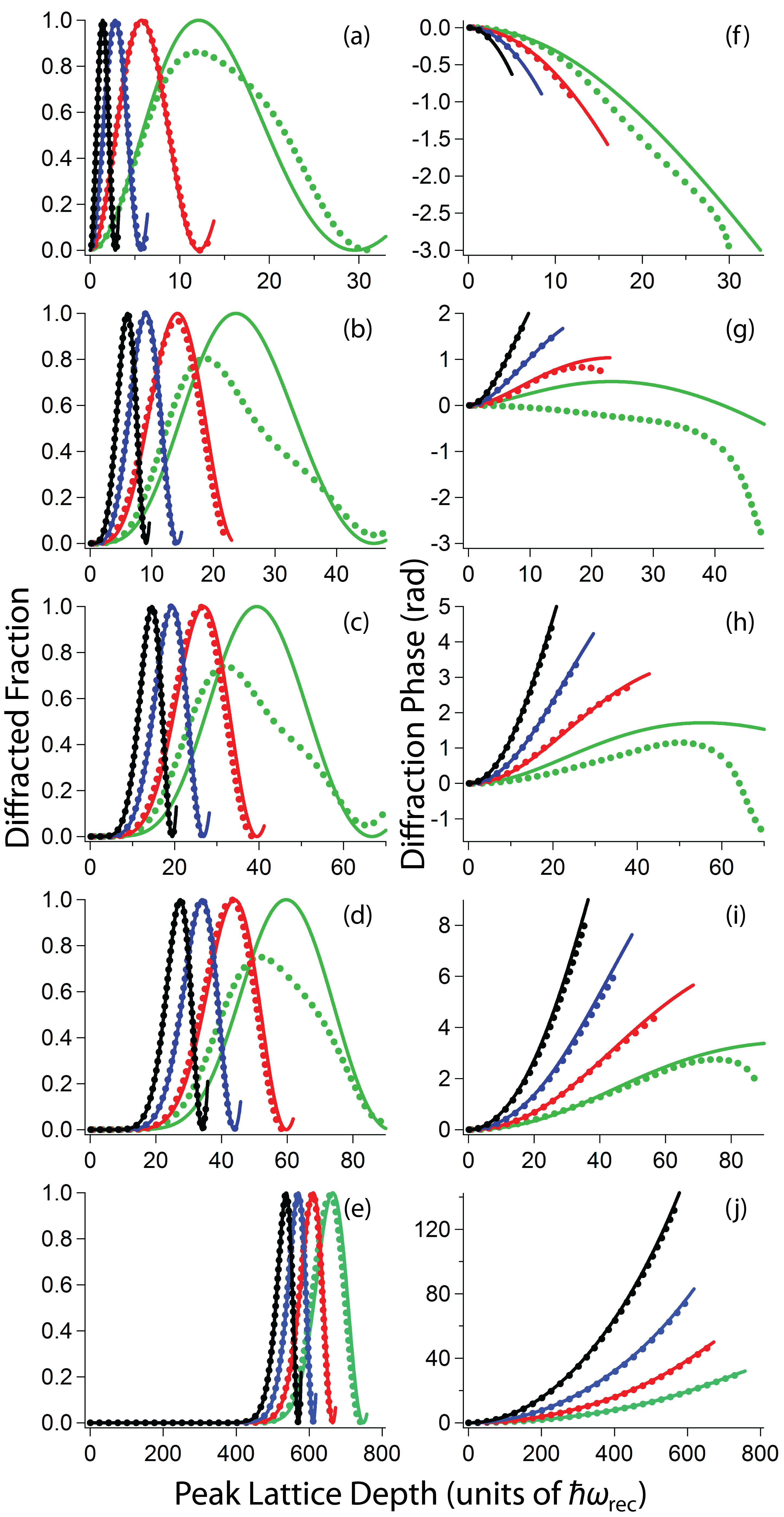}
	\caption{(Color online) (a-e) Diffracted fraction and (f-j) diffraction phase for  $N_B=1,2,3,4,16$ Bragg processes. Each graph contains results for Gaussian pulses with $4 \omrec \tau_{\rm 1/e}=10.1, 5.04, 2.52, 1.26$. Solid lines show the result of the Bloch-bands approach while the thick dashed lines show that of a full numerical integration in the free-particle basis. The two approaches maintain good agreement within the quasi-adiabatic regime (see text). We note that the phase shift is common to the initial and the final states.}
	\label{fig:Gochnauer_BandPicture_Figure6}
\end{figure}

We perform this test by comparing the results for a Bragg $\pi$-pulse evaluated by the Bloch-bands picture and a numerical integration of the problem in the free particle basis, as described in earlier work \cite{jami14} and summarized below. We use Gaussian pulses with different $1/e$ radii and observe (see Fig.\ref{fig:Gochnauer_BandPicture_Figure6}) that the two calculations agree in both population and phase for timescales comfortably satisfying the adiabaticty criterion as well as into the quasi-adiabatic regime convenient for and usually utilized in experiments. The signature of breakdown of the Bloch-bands picture is simultaneous with the observation of reduced diffraction efficiency (see green curves corresponding to the shortest pulses for $N_B=$ 1 to 4 in Fig.\ref{fig:Gochnauer_BandPicture_Figure6}) - a straightforward experimental diagnostic - making the Bloch-bands picture ideally suited for light pulse atom diffraction and interferometry. As expected, adiabaticity is restored for larger $N_B$ with the peak diffracted fraction for the shortest pulses calculated by numerical integration in the free-particle basis (dashed green lines) reaching 0.92, 0.97, 0.99 for $N_B=8,12,16$ respectively (the $N_B=16$ case is shown in Fig.\ref{fig:Gochnauer_BandPicture_Figure6}(e,j)).
	
\section{Full Numerical Integration}
	
For the full numerical simulation, we project onto a momentum-space basis $\{\dots,\ket{2\hbar k,g},\ket{\hbar k,e},\ket{0\hbar k,g},\ket{-\hbar k,e},\ket{-2\hbar k,g}\dots \}$, where the minimum and maximum momenta are chosen by testing for convergence. For the purposes of this discussion, we will truncate to five states \cite{foottruncstates}. While this basis contains both ground and electronically excited states, we find that the results do not depend on the detuning, $\Delta$, of the diffraction beams from the ground to excited state resonance, so long as the same two-photon Rabi frequency ($\omega_R^2/(2\Delta)$) is used. Utilizing the rotating frame approximation and transforming to the dressed-state basis gives a Hamiltonian of the form
\begin{equation*}
	\begin{pmatrix}
	4\hbar \omega_{\rm rec} & \frac{\hbar \omega_{\rm R}}{2} & 0 & 0 & 0\\
	\frac{\hbar \omega_{\rm R}}{2} & \hbar \omega_{\rm rec}-\hbar \Delta & \frac{\hbar \omega_{\rm R}}{2}  & 0 & 0 \\
	0 & \frac{\hbar \omega_{\rm R}}{2} & 0 & \frac{\hbar \omega_{\rm R}}{2} & 0 \\
	0 & 0 & \frac{\hbar \omega_{\rm R}}{2} & \hbar \omega_{\rm rec} - \hbar \Delta & \frac{\hbar \omega_{\rm R}}{2}\\
	0 & 0 & 0 & \frac{\hbar \omega_{\rm R}}{2}  & 4\hbar \omega_{\rm rec} 
	\end{pmatrix}.
	\end{equation*}
	
We numerically integrate to obtain the time evolution operator $\hat{U}$ for the full diffraction pulse and then extract the diffraction phase as the phase of the transition matrix element between two states of opposite momentum. For example, to simulate an $N=2$ pulse, the phase of the matrix element $\bra{2\hbar k,g} \hat{U} \ket{-2\hbar k,g}$ is compared to the phase of $\bra{0\hbar k,g} \hat{U} \ket{0\hbar k,g}$, as would be the case in an interferometer. For extracting the diffraction phase of a single state we compare to a much higher momentum state (e.g., $16\hbar k$ higher momentum), which should have negligible diffraction phase, and then confirm that neighboring high momentum states give the same results. This removes effects from momentum-independent AC stark shifts.

\section{Perturbative Scalings for Diffraction Phase and Deep Lattices}

\begin{figure}[b]
	\center
	\includegraphics[width=0.5\textwidth]{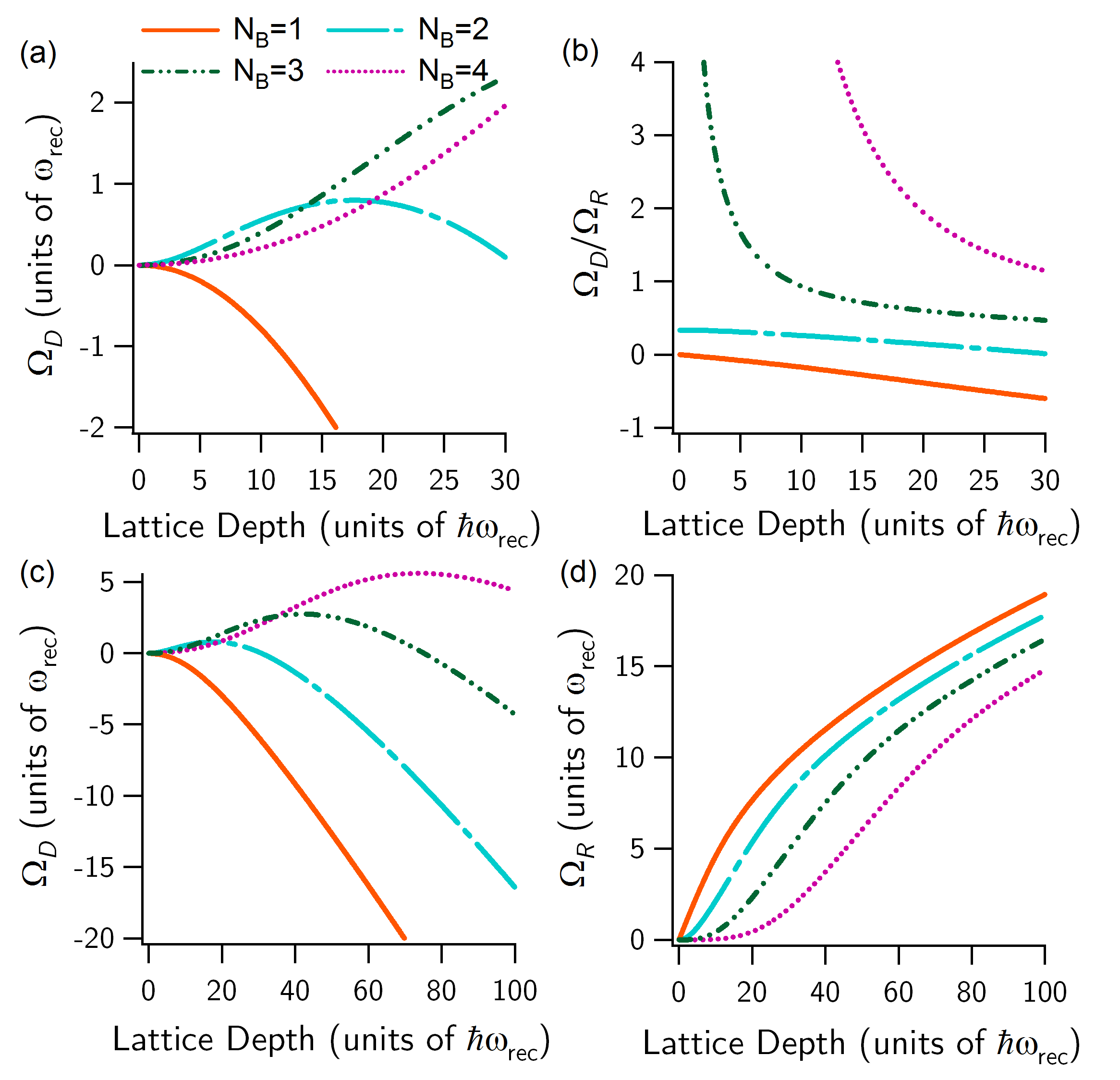}
	\caption{(Color online) (a) The frequency $\Omega_D$ associated with the diffraction phase plotted as a function of lattice depth for the four lowest level crossings in the band picture corresponding to the four lowest Bragg diffraction orders. (b) shows that the corresponding variation of $\Omega_D/\Omega_R$. (c) and (d) show the behavior of $\Omega_D$ and $\Omega_R$ respectively for large lattice depths.} 
	\label{fig:Gochnauer_BandPicture_Figure7}
\end{figure}

The behavior of the diffraction phase can be straightforwardly determined analytically in the limit of low lattice depth. Here, $\Omega_R$ can be taken to be the RHS of Eqn.(1) in the main text, which is inversely proportional to the $\pi-$pulse time. $\Omega_D$ is determined in second-order perturbation by the {\it first} (two-photon) off-resonant transition. For $N_B>1$,  $\Omega_D \propto (\frac{\omega_R^2}{\Delta})^2\frac{1}{N_B\omrec} = \frac{(U_0/\hbar)^2}{N_B\omrec}$. For the simple case of square pulses, $\Phi_D \propto \frac{\Omega_D}{\Omega_R}$, and this proportionality also indicates the behavior of $\Phi_D$ for a general pulse shape, with appropriate integration over the temporal profile of the diffraction pulse. Thus, $\Phi_D \propto \frac{\Omega_D}{\Omega_R} \propto \frac{1}{N_B}(\frac{U_0/\hbar}{\omrec})^{2-N_B}$. $\Omega_D$ is negative for $N_B=1$, and $\Phi_D \propto -\frac{1}{N_B}(\frac{U_0/\hbar}{\omrec})^{2-N_B}$. This behavior can be seen in the weak lattice limit of the curves in Fig.\ref{fig:Gochnauer_BandPicture_Figure7}(a,b). 

At larger lattice depth, the behaviour deviates signficantly from the perturbative expressions with $\Omega_D$ exhibiting a local maximum for all Bragg processes except $N_B=1$ (see Fig.\ref{fig:Gochnauer_BandPicture_Figure7}(c)). The locations of these maxima are in the diabatic regime for Bragg $\pi$-pulses. 
	
The deep lattice behavior of the Rabi frequencies or band gaps is shown in Fig.\ref{fig:Gochnauer_BandPicture_Figure7}(d). We note that in the limit of large lattice depth, the Rabi frequencies of all Bragg orders approach the values for the band gaps in the tight-binding limit corresponding to the harmonic oscillator spacing of $2\sqrt{U_0 \omrec/\hbar}$.

\begin{thebibliography}{34}
\bibitem[{foo()}]{footrabi}
	\bibinfo{journal}{When $q=0$, the outer paths are in a symmetric or
		anti-symmetric superposition of the $7^{\rm th}$ and $8^{\rm th}$ bands. In
		this case, an 8th order Bragg resonance condition is met for the outer paths.
		However, $\Omega_R$ and $\Omega_D$ are very small for this $N_B=8$ process
		making any population transfer or diffraction phase contributions negligible.
		Even at the largest values shown of 10 radians at 34 recoils in Fig.5(b), the
		population transfer for $N_B=8$ is $< 10^{-4}$ and the diffraction phase from
		outer paths $<0.1\,$rad.}
\bibitem[{foo(2019)}]{foottruncstates}
	\bibinfo{journal}{In practice, we find 17 states to be necessary for sufficient
		accuracy in modeling our $2^{\rm nd}$ order Bragg mirror pulses.}
\end{thebibliography}
\end{document}